\documentstyle[12pt,twoside]{article}
\normalbaselines
\font\tbf = cmbx12

\begin{document}

\indent
\vskip 1cm
\centerline{\tbf CHERENKOV RADIATION IN A GRAVITATIONAL WAVE
BACKGROUND}
\vskip 0.7cm

\vskip 0.3cm
\centerline{\tbf Alexander B. Balakin\footnote{e-mail: dulkyn@mail.ru} }
\vskip 0.3cm
\centerline{\it Kazan State University, Kremlevskaya street 18, 
420008, Kazan,  Russia,}
\vskip 0.5cm
\centerline{\tbf Richard Kerner \footnote{e-mail: rk@ccr.jussieu.fr} }
\vskip 0.3cm
\centerline{\it LPTL, Universit\'e Pierre et Marie Curie,  }
\centerline{\it Tour 22, Bo\^{\i}te 142, 4 Place Jussieu, 75005 Paris, France,}
\vskip 0.5cm
\centerline{\tbf Jos\'e P. S. Lemos\footnote{e-mail: lemos@kelvin.ist.utl.pt} }
\vskip 0.3cm
\centerline{\it CENTRA, Departamento de F\'{\i}sica, Instituto Superior 
T\'ecnico,}
\centerline{\it Av. Rovisco Pais 1, 1049-001 Lisboa, Portugal,}
\centerline{\it and}
\centerline{\it Observat\'orio Nacional - MCT,}
\centerline{\it Rua General Jos\'e Cristino 77, 20921, 
Rio de Janeiro, Brazil.}

\vskip 2cm
{\tbf Abstract} 
\indent 
{\small
A covariant criterion for the emission of 
Cherenkov radiation in the field of
a non-linear gravitational wave is considered in the framework of exact
integrable models of particle dynamics and electromagnetic wave propagation.
It is shown that  vacuum interacting with curvature can give rise to 
Cherenkov radiation. The conically shaped spatial distribution of radiation
is derived and its basic properties are discussed.}

\newpage

\section{Introduction}

The phenomenon of emission of radiation stimulated by a particle moving
with superluminal speed in a medium, first observed by Cherenkov and
Vavilov, and theoretically explained by Tamm and Frank, is one of the
cornerstones of classical electrodynamics (see, e.g., \cite{landau1} and
\cite{ginzburg}). This phenomenon, called Cherenkov radiation, is the
subject of many studies related to several fields of interest, from
theoretically motivated ones to collider detectors in accelerators,
for instance.
Here we are interested in studying the production of the Cherenkov radiation
in a gravitational field.

This problem can be studied through a model containing matter, the
gravitational field created by this matter, test relativistic particles
and  test photons induced by the moving particles.
The model can be characterized by a nonzero stress-energy
tensor in the right-hand side of the Einstein equations, which implies 
a nonzero Ricci tensor, and the Cherenkov photons may travel through the
material medium. This model was taken up in \cite{gupta}, where 
Cherenkov radiation was indeed predicted to be possible.

We would like to consider a second model, which contains an external
gravitational field, test relativistic particles and test photons.
In this model there is no material medium.
Due to the external gravitational field, there is a non-vanishing
Riemann curvature tensor in the background.

Now, it is known that the gravitational field itself can be considered as a sort of
medium. This possibility is suggested by a sentence of the book ``Classical
Field Theory'' of Landau and Lifshitz \cite{landau2}, in the problem of
paragraph 90, where it is stated that the gravitational field plays a role
of a medium with electric and magnetic permeabilities equal to $1/\sqrt{h}$,
where $h$ is the determinant of the three-dimensional spatial induced metric. 
Such analogy, of considering the gravitational field in vacuum as a medium, is 
useful in several instances.  If the gravitational field acts as a sort of medium,
one can then study the modifications induced on the speed of light in the
presence of the gravitational field itself. One can try to find  models,
in which a massive relativistic particle is able to move faster than photons.
If such models exist, then following the classical theory, one can expect
that the Cherenkov effect should become possible. Thus, one can use the
Cherenkov effect as a test for verifying models of interaction between the
electromagnetic and gravitational fields.

If we restrict ourselves to the so-called minimal coupling
between electro\-magnetic and gravitational fields 
(the covariant derivatives are based on the Christoffel
symbols), we are dealing with pure vacuum.
In this case the worldlines of photons are 
null geodesics, therefore, the velocity of the photon
coincides with the universal constant $c$. Thus, for pure vacuum
in a gravitational background one can expect that
there is no possibility for the emission of Cherenkov radiation.

The introduction of a non-minimal coupling between the electromagnetic
and gravitational fields in vacuum (the Lagrangian contains the
Riemann tensor) leads to models which can be classified as vacuum
interacting with curvature. An example of such a vacuum has been
proposed by Drummond and Hathrell \cite{drummond}. This example has
appeared as a result of one-loop calculations in the framework of
Quantum Electrodynamics (QED) in a gravitational background and it was
shown that the velocity of light can differ from $c$. The vacuum of
Drummond and Hathrell is in a class of vacua called ``non-trivial QED
vacua'' \cite{dittrich} or ``modified QED vacua'' \cite{scharnhorst}.
A general property of a non-trivial QED vacuum is that the velocity of
light differs from $c$. In the other words, a non-trivial QED vacuum
behaves as a sort of medium, a quasi-medium, with effective refractive
index $n \neq 1$.  Thus, it is natural to ask the following question:
is the Cherenkov effect possible in a vacuum interacting with a
gravitational field?  In this paper we want to consider the Cherenkov
radiation in a gravitational wave (GW) background with vacuum
interacting with curvature.  Gravitational radiation is an example of
a non-stationary field. Thus this quasi-medium (of a vacuum
interacting with curvature) behaves as a nonstationary medium. As a
result, as we will see, the effective refractive index depends on
time.

The purpose of our paper is to show that a covariant
analysis of this problem, based on exact solutions for the coupled system
of a charged particle and an electromagnetic wave in a nonlinear
gravitational wave background, enables one to treat it exactly in some cases. 
The results of our investigation show that a GW background field, 
quasi-media and true material media display the
Cherenkov effect, while pure vacuum deos not.

The paper is organized as follows.  In Section 2, starting from the
classical description of the Cherenkov effect in the material medium
with constant refractive index, we formulate a covariant criterion and
the necessary condition for the emission of Cherenkov radiation. In
Section 3 we integrate exactly the equations of particle motion as
well as the Maxwell equations in the GW background and apply the
criterion and the necessary condition for Cherenkov radiation. In
Section 4 we consider the spatial properties of Cherenkov radiation by
using two examples of longitudinal and transversal particle motion. In
the Appendices we obtain and briefly discuss the details of the exact
solutions of Maxwell equations.

\section{The Cherenkov radiation in a curved space-time:
the general criterion}

The Cherenkov radiation can be explained in simple terms in Minkowskian
space-time (for details, see \cite{landau1} and \cite{ginzburg}). Consider
a charged particle moving uniformly with velocity ${\bf V}$ in a static
isotropic dielectric medium with the refraction index $n$. 

This particle induces an electromagnetic field which has the following 
Fourier representation:
\begin{equation}
{\bf A}(t, {\bf r}) = {\displaystyle{\sum_{(l)}}} \, 
q_{(l)}(t) \  {\bf A}_{(l)}({\bf r}) + q^*_{(l)}(t) \ {\bf A}_{(l)}^*({\bf r})
\,,
\label{decom}
\end{equation}
where ${\bf A}(t, {\bf r})$ is the potential three-vector, $q_{(l)}(t)$ are
the Fourier amplitudes with $(l) = 1,2,....\infty$. The modes
${\bf A}_{(l)}({\bf r})$ are given by
\begin{equation}
{\bf A}_{(l)}({\bf r}) = \sqrt{4 \pi} \ \frac{c}{n} \, 
{\bf e}_{(l)} \, e^{(i {\bf k}_{(l)} \cdot {\bf r})} \,,
\label{fa}
\end{equation}
where ${\bf k}_{(l)}$ with the fixed $(l)$ is the wave three-vector of the mode
enumerated by $(l)$, and ${\bf e}_{(l)}$ is the polarization three-vector with 
unit length. Then, in the Lorentz gauge $\partial_{\mu} A^{\mu} = 0$, and from 
Maxwell's equations \cite{ginzburg}
\begin{equation}
\frac{n^2}{c^2} \, \frac{\partial^2 \, {\bf A}}{\partial t^2} -
\bigtriangleup {\bf A} = \frac{4\pi}{c} \, {\bf j} \, , \ \ \, \ \ {\rm and} \,
\ \ \,  \frac{n^2}{c^2} \, \frac{\partial^2 \, {\varphi}}{\partial t^2} -
\bigtriangleup {\varphi} = 4\pi \, \rho \, ,
\label{threemax}
\end{equation}
we get, in terms of Fourier amplitudes, the following equations
\begin{equation}
\frac{d^2 q_{(l)}}{dt^2} + \omega^2_{(l)} \, q_{(l)} = \frac{1}{c} \, \int \, 
{\bf j}(t,{\bf r}) \, {\bf A}^*_{(l)}({\bf r})  \, d^3 {\bf r}  \,,
\label{modeseq}
\end{equation}
where $ \omega_{(l)} \equiv c k_{(l)} / n$.
For a point particle with electric charge $e$ moving with
constant velocity $\bf V$, the electric current ${\bf j}(t,{\bf r})$ has the 
form
\begin{equation}
{\bf j}(t,{\bf r}) = e \, {\bf V} \, \delta \, ( {\bf r} - {\bf V} t) \,,
\label{current}
\end{equation}
and we obtain from (\ref{modeseq}) the equation of an oscillator solicited
by an external force:
\begin{equation}
\frac{d^2 q_{(l)}}{dt^2} + \omega^2_{(l)} \, q_{(l)} =  
\sqrt{4 \pi} \ \frac{e}{n} \, 
({\bf e}_{(l)} \cdot {\bf V} ) \, e^{- i ({\bf k}_{(l)} \cdot {\bf V}) t } \,.
\label{oscillator}
\end{equation}
The external force in the right hand side of the equation (\ref{oscillator})
is periodic with the frequency $\Omega_{(l)} = ({\bf k}_{(l)} \cdot {\bf V}) = 
k_{(l)} V \cos{\theta_{(l)}}$, where $\theta_{(l)}$ is
the angle between the direction of the wave three-vector ${\bf k}_{(l)}$ and
the three-velocity of the particle ${\bf V}$. 
When 
\begin{equation}
\Omega_{(l)} = \omega_{(l)}, \quad 
{\rm or \ \ equivalently} \quad
V \cos{\theta_{(l)}}= c/n \,,
\label{reso}
\end{equation}
a resonance occurs, making the oscillator's amplitude $q_{(l)}$ grow linearly
with $t$. This resonance condition is interpreted in the book \cite{ginzburg}
as a condition for the existence of Cherenkov's radiation. 
Since $\vert \cos{\theta_{(l)}} \vert \leq 1$, one sees from (\ref{reso}) that
the resonance is formally possible when $V \geq c/n$. Usually one excludes the 
case $\cos{\theta_{(l)}} = 1$ which corresponds to the radiation with
zero aperture, and one obtains that the condition for the Cherenkov 
radiation emission is 
\begin{equation}
V>\frac{c}{n}\,,
\label{velocitycondition1}
\end{equation}
i.e., the velocity of the particle should be greater than the velocity 
of light in the medium $c/n$. It is also clear that for fixed $V$ the 
maximal angle of radiation, the Cherenkov angle, is given by the condition
\begin{equation}
\cos\theta_0=\frac{c}{nV}\,. 
\label{angle1}
\end{equation}
In order for this radiation to represent a real electromagnetic wave
propagating in a medium with the refraction index $n$, the
relation between the electromagnetic wave frequency $\omega$ and the modulus 
of the wave vector $K$ should be $\omega = Kc/n$ \cite{landau1,ginzburg}.
Considering the identifications $K \equiv k_{(l)}$ and 
$\theta \equiv \theta_{(l)}$, where the fixed $(l)$ 
corresponds to the resonant mode, we obtain the following criterion from
(\ref{reso}) :
\begin{equation}
\omega \equiv K \frac{c}{n} \  = 
K V \cos{\theta} = \  ({\bf K} \cdot {\bf V})  \,.
\label{okv}
\end{equation}
This criterion for Cherenkov's radiation emitted by a charged particle must
be re-formulated in a covariant way when the particle is supposed to move
in a curved spacetime.  In covariant formulation we cannot use the
three-velocity vector ${\bf V}$, or the Cherenkov angle $\theta_0$, because
they are not Lorentz invariants.  There are two covariant vectors in this
problem, the time-like momentum four-vector of the charged particle $P^i$,
normalized according to $P_iP^i = m^2 c^2 > 0$, and the wave four-vector
$K_i$ characterizing the electromagnetic plane wave which can, in principle,
propagate inside the medium in a given space-time.  We say therefore that the
Cherenkov radiation can exist when the following equality is satisfied :
\begin{equation}
K_m P^m = 0\,,
\label{criterion}
\end{equation}
where latin indices are space-time indices running from 0 to 3, 
and where we use the metric with signature $(+\,-\,-\,-)$.
 
One can easily recover the relations (\ref{velocitycondition1})-(\ref{okv}). 
In a standard three-dimensional context in Minkowskian space-time the
relationship (\ref{criterion}) can be rewritten in the usual form 
\begin{equation}
K_m P^m = K_0 P^0 - {\bf K} \cdot {\bf P} =
\frac{P^0}{c} \left( \omega - {\bf K}\cdot {\bf V} \right) =
\frac{P^0 \omega}{c} \left( 1- \frac{1}{c} {\bf n} \cdot {\bf V} \right) = 0\,.
\label{3criterion}
\end{equation}
The classical definition of frequency is $\omega =cK^0$, the vector refraction
index is ${\bf n}=c {\bf K}/\omega$, and its square is $n^2 = {\bf n}^2$.
We can now reproduce the criterion to observe the Cherenkov radiation in a
flat space-time, i.e., we can reproduce equations
(\ref{velocitycondition1})-(\ref{okv}). For instance, using the 
definition of the angle $\theta$, i.e., 
$\cos \theta =  {\bf K} \cdot {\bf V}/ K V = {\bf n}\cdot {\bf V}/ nV$,
as well as the condition $\vert \cos \theta \vert \leq 1$, one can see from
(\ref{3criterion}) that $V  \geq c/ n$, i.e., particle's three-velocity
should be bigger than speed of light in the medium, recovering
(\ref{velocitycondition1}).
Equation (\ref{angle1}) and (\ref{okv}) also follow in a 
straightforward way from (\ref{3criterion}).

Let us now find the necessary condition for the existence of
Cherenkov's radiation. It can be given in three equivalent ways, the first
involving the square $K_i K^i$ of the wave four-vector $K^i$, the second one
involving the square $n^2$ of the scalar refraction index, and the third
one involving the phase velocity of light in the medium, $v_{\rm ph}$. Note
that this necessary condition does not depend on particle's momentum $P^i$.
It is well known  that the frequency $\omega$ (considered as a scalar
quantity), and the four-vector $K_i^*$ (the spatial part of the wave
four-vector) may be defined by the following relations (see, e.g., 
\cite{syng}) :
\begin{equation}
\frac{\omega}{c} \equiv K_i U^i\,,
\label{omega}
\end{equation}
and
\begin{equation}
 K^{*}_i \equiv K_j \Delta^j_i\,.
\label{3wavevector}
\end{equation}
Here $U^i$ is the four-velocity vector of the medium (or of the observer,
if we consider propagation in pure vacuum), and $\Delta^l_i$ is the projector
defined as
\begin{equation}
\Delta^i_j \equiv \delta^i_j - U^i U_j\,.
\label{projector}
\end{equation}
Following the standard definition (see p.290 in \cite{landau1}), we introduce
now the four-vector refraction index $n_i$:
\begin{equation}
n_i = \frac{c}{\omega} K_i^*\,.
\label{vecopticalindex}
\end{equation}
The vector refraction index $n_i$ is in general a spacelike four-vector,
orthogonal to the four-velocity $U_i$, and only in the special case of 
isotropic medium it reduces to a scalar. Its absolute value is referred to as 
the scalar index of refraction, (or refraction index, for simplicity). Its 
square is by definition given by
\begin{equation}
n^2 \equiv - g^{ik} n_i n_k \,.
\label{scalaropticalindex}
\end{equation}
We can now write the following useful identity satisfied by $K^i$, 
\begin{equation}
K_i = U_i (K_l U^l) + K^{*}_i\,.
\label{kdecomposition}
\end{equation}
Now, in order to use the criterion (\ref{criterion}) we note that $P^i$ and
$K^i$ must be orthogonal. This happens only when  $K_i$ is spacelike, since
the $P^i$ four-vector is timelike.
 From the equation (\ref{kdecomposition}) we obtain the square of $K_i$, 
\begin{equation}
K^2 \equiv -g^{im}K_i K_m = - K_m K^m= - (K^m U_m)^2 - K^l K^s \Delta_{ls}\,.
\label{squareK}
\end{equation}
This equation can be written explicitly in two ways, 
\begin{equation}
K^2= \left(\frac{\omega}{c}\right)^2 (n^2-1)\,,
\label{ksquare}
\end{equation}
and 
\begin{equation}
K^2= (K^*)^2 - \frac{\omega^2}{c^2} \,,
\label{signksquare}
\end{equation}
where 
\begin{equation}
(K^*)^2 \equiv - K^l K^s \Delta_{ls} \,.
\label{kstarsquare}
\end{equation}
We see from the equation (\ref{ksquare}) that the sign of $K^2$
coincides with the sign of $(n^2-1)$. This means that  $K_i$ is spacelike
for $n^2>1$. From the equation (\ref{signksquare}) it can be inferred that
$K_i$ is spacelike when $\omega^2/c^2<(K^*)^2$, i.e. when the phase velocity
of light $v_{\rm ph}\equiv\omega/ K^* $ obeys $v_{\rm ph}<c$.
We can use one of the three equivalent invariant forms of the necessary
condition for existence of Cherenkov radiation, namely,
\begin{equation}
K_m K^m < 0\,, \quad {\rm or,} \quad n^2>1\,, \quad {\rm or,} \quad 
v_{\rm ph}<c\,.
\label{necessary}
\end{equation}  
All of them require the knowledge of the four-vector $K^i$, which 
is obtained from the corresponding solution of Maxwell equations.

In order to use the criterion (\ref{criterion}) in a curved background we
have to resort now to a specific model. We use as a background the pp-wave
solution of Einstein equations in vacuum. We can then determine 
the four-momentum $P^i$ of
a particle moving in this background, and find a specific solution
of Maxwell equations, in the same gravitational background. After
that, we can examine the criteria of existence of the Cherenkov radiation,
and establish its spatial properties in this background. In a curved
space-time different spatial directions are generally non-equivalent, which
implies that one should know the evolution of particle's four-momentum $P^i$
with arbitrary initial data in order to be able to use explicitly the
criterion (\ref{criterion}).

\section{Charged particle in the GW background and the Cherenkov radiation}

\subsection{The gravitational wave background}

Let us consider the space-time described by the exact pp-wave solution of
Einstein's equations in vacuum \cite{kramer}. The metric describing a
gravitational wave propagating in the $x^1$ direction is supposed to take on
the following form:
\begin{equation}
ds^{2} =  2 du dv -
L^{2} \left[e^{2\beta}(dx^{2})^{2} + e^{-2\beta}(dx^{3})^{2} \right],
\label{metric}
\end{equation}
where
\begin{equation}
u =  \frac{ct-x^{1}}{\sqrt{2}}, \quad  v = \frac{ct+x^{1}}{\sqrt{2}}
\label{times}
\end{equation}
are the retarded and the advanced times, respectively. The functions
$ L$ and $ \beta$ depend only on the variable $u$, $L=L(u)$ and
$\beta=\beta(u)$.

The  pp-wave metric (\ref{metric}) is invariant under the $G_{5}$ symmetry
group and admits the following set of five Killing vector fields $\xi_{(r)}$ 
(where the index $(r)$ takes on the values $(v), (2), (3), (4)$ and $(5)$, and
characterizes each vector), 
$$ \xi^{i}_{(v)}= \delta ^{i}_{v} \,,  \, \ \  \, \ \ 
\xi^{i}_{(2)}= \delta ^{i}_{2} \,, \, \ \ \, \ \ 
\xi^{i}_{(3)}= \delta ^{i}_{3} \,, $$
\begin{equation}
\xi^{i}_{(4)}= x^{2}  \delta^{i}_{v}  -  \delta^{i}_{2} 
\int{g^{22}(u)du} \,, \quad
\xi^{i}_{(5)} = x^{3}  \delta^{i}_{v}
-  \delta^{i}_{3}  \int{g^{33}(u)du} \,. 
\label{killings}
\end{equation}
Here $g^{\alpha \beta}(u)$ ($\alpha, \beta = 2,3$) are the
contravariant components of the metric tensor. The vector
$\xi^{i}_{(v)}$ is isotropic, covariantly constant and orthogonal to
the other four ones, 
\begin{equation}
\nabla_{k} \  \xi^{i}_{(v)}=0, \quad g_{ik}  \
\xi^{i}_{(v)} \  \xi^{k}_{(r)} =0.
\label{ortho}
\end{equation}
The three vectors
$\xi^{i}_{(v)}, \xi^{i}_{(2)}, \xi^{i}_{(3)}$
form the Abelian subgroup $G_{3}$.
The two functions  $L(u), \beta  (u)$  are coupled
by the Einstein equation, unique in this case,
\begin{equation}
L^{''} + L \  (\beta^{'})^{2} = 0\,.
\label{einstein}
\end{equation}
The function $\beta (u)$  can be chosen at will, and once given, one may
solve the equation (\ref{einstein}) for $L(u)$. The curvature tensor has two
non-vanishing components:
\begin{equation}
- R^{2}_{\cdot u2u} = R^{3}_{\cdot u3u} =
L^{-2} \left[L^{2} \  \beta^{'} \right]^{'} \,. 
\label{riemann}
\end{equation}
Both the Ricci tensor $R_{ik}$ and the curvature scalar $R$ are equal to zero.

\subsection{Particle dynamics in the GW background}

Solving the geodesic equation for a  particle with mass $m$ in the GW 
field (\ref{metric})
\begin{equation}
\frac{D P_i}{D\tau} = 0 \,, \quad {\rm with\,} \quad
P^i = mc \frac{d x^i}{d\tau} \,,
\label{eqmotion}
\end{equation}
and using the well-known property of the Killing vectors (\ref{killings})
\cite{mtw}, we obtain the following expressions for the components of the
momentum $P_i$ :
\begin{eqnarray}
&
P_v \equiv P_i \xi^i_{(v)} = {\rm const} \equiv  C_v\,, \label{momentum1}
\\&
P_\alpha \equiv P_i \xi^i_{(\alpha)} = {\rm const} \equiv C_\alpha\,, &
\label{momentum2}
\\
&
P_u = \frac{1}{2 C_v}\left[ m^2 c^2 - g^{\alpha \beta} C_\alpha 
C_\beta \right]\,, & \label{momentum3} 
\label{pu}
\end{eqnarray}
where, the last equation for the  component $P_u$ of the 
momentum followed from the normalization condition.

Since $C_v$ and $C_\alpha$ are constants, they also represent the initial
values of the corresponding momentum components at the initial surface
defined by $u = 0$. These data determine the character of the particle's motion.
For example, when $C_\alpha = 0$, i.e., when the particle moves initially
along the longitudinal direction (the direction of propagation of the GW
$x^1$), then it will always move along that direction without acceleration.
When $C_\alpha \neq 0$, the dynamical effects on the particle induced by the
GW appear in its longitudinal motion (see the expression for $P_u$ (\ref{pu})).
Thus, in the field of GW, the criterion (\ref{criterion}) yields the following
equation :
\begin{eqnarray}
& K_mP^m=K_uP_v + K_vP_u+ K_\alpha P^\alpha=\nonumber\\&
=K_u C_v + K_v  \frac{1}{2 C_v}\left[ m^2 c^2 - g^{\alpha \beta}(u)
C_\alpha C_\beta \right]+ g^{\alpha \beta}(u) K_\alpha C_\beta = 0\,.&
\label{gwcriterion}
\end{eqnarray}
We should now represent explicitly the components of the wave four-vector
$K_i$, and we also have to check the condition (\ref{necessary}) necessary
for the emission of the Cherenkov radiation.

\subsection{Maxwell's equations in the GW background}

\indent
The solutions of Maxwell's equations in a gravitational wave background have
been discussed in a strong GW field in the framework of geometrical optics
approximation (e.g., \cite{sbytov}), in the case of weak GW (see, e.g.
\cite{calura} and the references therein), and in the case of strong GW
without the geometrical optics approximation \cite{bala1}.

In this section we
follow the approach presented in the article \cite{bala1}, in which it has
been shown that the problem of propagation of electromagnetic waves in a GW
background (in pure vacuum, or in vacuum interacting with curvature) is an
exactly integrable problem. This approach has some advantages. For instance,
if we deal with the vacuum case, it is not needed to restrict the solution
either to the geometrical optics, or to the weak gravitational field approximations.
 
On the other hand, Maxwell's equations in a medium in presence of a
gravitational wave field pose some additional problems and lead to
a nonreducible system of equations. Therefore, when we deal with the
Maxwell equations in a spatially isotropic medium, we have to consider
the eikonal equation, i.e., we use the geometrical optics approximation
(see, e.g, \cite{syng} for the description of covariant formalism). 
In the Appendix A we describe the methods of obtaining the solutions of
Maxwell equations for pure vacuum, for vacuum interacting with curvature,
for spatially isotropic dielectric medium and for spatially isotropic medium
interacting with curvature.
Below, we extract the information given in the
Appendix A on the wave four-vector $K^i$, the dispersion relation
$\omega(K^*)$, and the refraction index, and use it for investigating the
possibility of existence of Cherenkov's radiation in a GW background for 
pure vacuum case and for vacuum interacting with curvature. In the Appendix B
we study briefly the possibility of Cherenkov's radiation in a spatially
isotropic medium interacting with curvature.  

\subsubsection{Pure vacuum}

As it is shown in the Appendix A (see section A1.1), the wave four-vector $K_i$ has
the following form:
\begin{equation}
K_{i}  = \delta^{u}_{i} \left[
- \frac{k_{\alpha}k_{\beta}}{2k_{v}}g^{\alpha \beta}(u) \right] +
\delta^{v}_{i} k_{v} + \delta^{2}_{i}k_{2} + \delta^{3}_{i} k_{3}\, ,
\label{wavevector10}
\end{equation}
where $ k_{v}, k_{2}, k_{3}$ are arbitrary constants, which, together with
the constant $k_{1}$, give the initial components of the wave four-vector. 
One can check easily that the square of this vector is equal to zero, 
\begin{equation}
g^{ij} K_{i} K_{j} = 0.
\label{wavevectorsquare10}
\end{equation}
Thus, the wave four-vector is isotropic and the phase velocity of the 
electromagnetic wave is equal to the speed of light in vacuum, yielding 
\begin{equation}
\omega = c K^* \, .
\label{omega1}
\end{equation}
 From the equations (\ref{ksquare})-(\ref{signksquare}), 
the scalar refraction index is given by
\begin{equation}
n=1 \, .
\label{omega11}
\end{equation}
Therefore, the criterion (\ref{criterion}) (or \ref{gwcriterion}) can not be
satisfied and there is no possibility for the Cherenkov radiation to be
emitted in pure vacuum.
 
\subsubsection{Vacuum interacting with curvature}

As it is shown in section A1.2 of Appendix A, the interaction of vacuum with
curvature (studied first in \cite{drummond}) leads to birefringence, i.e.,
the wave four-vector $K_i$  depends on polarization. From the Appendix A we
have that there are two distinct electromagnetic waves, each one represented
by a corresponding wave four-vector, $K_i^{(2)}$ or $K_i^{(3)}$, given by  
\begin{equation}
K^{(2)}_{i}= \delta^{u}_{i} \left[
- \frac{k_{\alpha}k_{\beta}}{2k_{v}}g^{\alpha \beta}(u)
+ k_{v} q R^{3}_{\cdot u3u} \right] + \delta^{v}_{i} k_{v} +
\delta^{2}_{i}k_{2} + \delta^{3}_{i} k_{3}\, ,
\label{tidalk20}
\end{equation}
\begin{equation}
K^{(3)}_{i} = \delta^{u}_{i} \left[
- \frac{k_{\alpha}k_{\beta}}{2k_{v}}g^{\alpha \beta}(u)
- k_{v} q R^{3}_{\cdot u3u} \right] + \delta^{v}_{i} k_{v} +
\delta^{2}_{i}k_{2} + \delta^{3}_{i} k_{3}\,, 
\label{tidalk30}
\end{equation}
where $ k_{v}, k_{2}, k_{3}$ are again arbitrary constants, which, 
with $k_{1}$, give the initial components of the wave four-vector. 
Introducing the term $s^{(A)} \equiv (-1)^{(A)}$ with $(A)=2,3$ one can 
condense formulae (\ref{tidalk20})-(\ref{tidalk30}) into a single 
formula
\begin{equation}
K^{(A)}_{i}  = \delta^{u}_{i} \left[
- \frac{k_{\alpha}k_{\beta}}{2k_{v}}g^{\alpha \beta}(u)
+s^{(A)} k_{v} q R^{3}_{\cdot u3u} \right] + \delta^{v}_{i} k_{v} +
\delta^{2}_{i}k_{2} + \delta^{3}_{i} k_{3}.
\label{tidalk230}
\end{equation}
The vectors $K^{(2)}_{i}$ and $K^{(3)}_{i}$ are orthogonal to each other, 
\begin{equation}
g^{ij} K^{(2)}_{i} K^{(3)}_{j} = 0,
\label{k2k3ortho0}
\end{equation}
and their squares are given by,
\begin{equation}
g^{ij} K^{(2)}_{i} K^{(2)}_{j}
= 2 q (k_{v})^2 R^{3}_{\cdot u3u}\;, \quad
g^{ij} K^{(3)}_{i}K^{(3)}_{j}
= - 2 q (k_{v})^2 R^{3}_{\cdot u3u}.
\label{k2k3square}
\end{equation}
Here $ R^{3}_{\cdot u3u}$ is the independent component of the Riemann
tensor and $q$ is the interaction parameter of the vacuum with
curvature (see Appendix A).  Note that the signs in
(\ref{k2k3square}), depend on the sign of $q R^{3}_{\cdot u3u}$, this
being positive or negative at different times and in different spatial
points. Since each of the two vectors $K^{(A)}$, has opposite sign, when
$R^{3}_{\cdot u3u} \neq 0$, one of the two vectors has a negative
square. Thus, one of the electromagnetic waves propagates with phase
velocity less than the velocity of light in this vacuum interacting with
curvature, and the necessary condition for Cherenkov radiation
(\ref{necessary}) is satisfied.

The frequency and the spatial part of the wave vector depend on the index 
$(A)$. Thus, using $U_i = \frac{1}{\sqrt2}(\delta_i^u + \delta_i^v)$
for the four-velocity of the medium (or observer) at rest, 
one can write from (\ref{tidalk230}) the formula for the two frequencies
\begin{equation}
\frac{\omega^{(A)}}{c} \equiv K^{(A)}_i U^i = \frac{k_v}{\sqrt{2}}
\left( 1 + s^{(A)} q R^3_{\cdot u3u} -
\frac{1}{2k^2_v} g^{\alpha \beta} k_\alpha k_\beta \right).
\label{aomegasolution}
\end{equation}
In a spherical representation of the initial components of the wave 
four-vector we can write, 
\begin{equation}
k_1 = - k \cos\theta_0\,, \ \
k_2 = - k \sin\theta_0 \cos\varphi_0\,, \ \
k_3 = - k \sin\theta_0 \sin\varphi_0\,, \ \
k_v = \sqrt{2} k \sin^2{\frac{\theta_0}{2}}\,,
\label{angles0}
\end{equation}
where $\theta_0$ and $\varphi_0$ are the polar and azimuthal angles of the 
initial direction of the propagation of the electromagnetic wave. 
Note that $k_1^2 + k_2^2 + k_3^2 \equiv k^2$, and  
$k_v = \frac{1}{\sqrt2} (k_0 + k_1)$. 

Now, when $\theta_0 \neq 0$, the frequencies depend on retarded time $u$
as well as on $\theta_0$ and $\varphi_0$ and (\ref{aomegasolution}) yields
\begin{equation}
\frac{\omega^{(A)}}{c} = k \sin^2{\frac{\theta_0}{2}} \left[
1 + s^{(A)} q R^3_{\cdot u3u} + \frac{1}{L^2} \cot^2{\frac{\theta_0}{2}}
\left(\cosh{2\beta} - \cos{2\varphi_0} \sinh{2\beta} \right) \right] \,. 
\label{aomega1}
\end{equation}
Analogously, for the two scalar indices of refraction we obtain 
\begin{equation}
(n^{(A)})^2  = 1 - \frac{ 4 s^{(A)} q R^3_{\cdot u3u}}{\left[
1 + s^{(A)} q R^3_{\cdot u3u} + \frac{1}{L^2}\cot^2{\frac{\theta_0}{2}}
\left(\cosh{2\beta} - \cos{2\varphi_0} \sinh{2\beta} \right) \right]^2} \,.
\label{aindex0}
\end{equation}
Using (\ref{ksquare}), (\ref{k2k3square}) and (\ref{angles0}) one can 
describe explicitly the dispersion properties of the vacuum interacting with 
curvature, i.e., represent the refraction index $n^{(A)}$ as a function of 
$\omega^{(A)}$ and $k$: 
\begin{equation}
(n^{(A)})^2  = 1 - \left(\frac{ c \ k \ }{ \ \omega^{(A)}}\right)^2
4 s^{(A)}  q R^3_{\cdot u3u}  \sin^4{\frac{\theta_0}{2}} \,.
\label{noku}
\end{equation}
We see that the refraction index depends not only on $\omega^{(A)}$ and $k$,
but also on the polarization index $(A)$, on the polar angle $\theta_0$ and
on the retarded time $u$. Therefore, analysing the formula (\ref{noku})
following the lines of the book \cite{ginzburg} one can say that the vacuum
interacting with curvature behaves as a spatially anisotropic nonstationary
quasi-medium, admitting birefringence and transient radiation phenomena.

In the particular case $\theta_0 = \pi$ one obtains  $k^1 = - k$, i.e., the
GW and the electromagnetic wave propagate in opposite directions. For 
the index of refraction  we obtain the following simple formula :
\begin{equation}
n^{(A)} = \frac{1 -  s^{(A)} q R^3_{\cdot u3u}}
{1 +  s^{(A)} q R^3_{\cdot u3u}}\, .
\label{anpi}
\end{equation}
One of the two  indices of refraction  exceeds unity $n^{(A)} > 1$, when  
$s^{(A)} q R^3_{\cdot u3u} < 0 $. 
For a periodic gravitational radiation the term $R^3_{\cdot u3u}$ oscillates, 
so, the indices of refraction  $n^{(A)}$ become periodically larger or smaller
than unity.  The wave four-vector periodically becomes space-like or
time-like and the electromagnetic wave propagates with subluminous or
supraluminous phase velocity, respectively. Thus, for one of the two 
orthogonally polarized waves (for $(A)=2$ or $(A)=3$) the necessary condition
for the Cherenkov radiation emission is satisfied.

In the limiting case  $\theta_0 = 0$, one obtains $k_v = 0$. The wave
vectors become isotropic and both of the refraction indices are equal to
one, $n^{(A)} \equiv 1$. There is no Cherenkov radiation in this direction.

\section{Spatial properties of the Cherenkov radiation in a GW background}

Landau et al. give a problem for finding the conditions for Cherenkov
radiation, as well as, the cone of its wave vectors for a particle
moving uniformly in a uniaxial non-magnetic crystal, along the optical
axis, and at right angles to it (\cite{landau1}, p.407).  In such a
crystal there is birefringence, i.e., the wave four-vector $K_i$ and
the velocity of the wave depend on the polarization. Usually, there
are two waves, the ordinary and the extraordinary, propagating with
different speeds.  In this problem of Landau et al., it is found that
along the optical axis the radiation is contained in a circular cone,
whereas at right angles to it there are two cones, corresponding to
the ordinary and extraordinary waves, the one for the ordinary wave
being circular, and the other for the extraordinary wave being a cone
with different geometry, a non-circular one.

We will see now that in our problem there are two privileged directions 
along which one can find the geometry of the cones of the Cherenkov radiation.
Indeed, the GW background defines the direction of propagation of the wave,
the longitudinal direction, and the directions orthogonal to it, i.e. the
transversal ones.

In order to describe the spatial properties of Cherenkov radiation, 
we study the criterion (\ref{gwcriterion}) for 
the case of vacuum interacting with curvature (see section 3.3.2, 
for another case see Appendix B).  
For this model the exact solutions for the wave four-vectors, 
$K_i^{(2)}$ and $K_i^{(3)}$, are given in equations 
(\ref{tidalk20})-(\ref{tidalk30}). Then equation (\ref{gwcriterion}) 
has the form 
\begin{equation}
\left(- \frac{1}{2 k_v}g^{\alpha \beta} k_\alpha k_\beta +
k_v s^{(A)} q R^3_{\cdot u3u} \right) C_v + k_v \frac{1}{2 C_v}
\left(m^2 c^2 - g^{\alpha \beta} C_\alpha C_\beta \right)
+ g^{\alpha \beta} k_\alpha C_\beta = 0.
\label{dtidalcriterion}
\end{equation}
Let us now consider two cases of particle motion, one motion being
along the longitudinal direction, the other along one particular 
transversal direction.

\subsection{Longitudinal particle motion}

This case describes the situation of a particle moving in a direction
parallel to the $x^1$ axis along which the GW propagates.
In this case the initial transversal components of the particle momentum 
vanish, i.e., 
\begin{equation}
C_\alpha =0, 
\label{calpha}
\end{equation}
and the equation (\ref{dtidalcriterion}) is simplified,
\begin{equation}
\left(- \frac{1}{2 k_v}g^{\alpha \beta} k_\alpha k_\beta +
k_v s^{(A)} q R^3_{\cdot u3u} \right) C_v + k_v \frac{m^2c^2}{2 C_v} = 0\,.
\label{tc1}
\end{equation}
 From the definition $P^1=\frac{1}{\sqrt2}(P_u-P_v)$, from 
the equation $P^1=mV/\sqrt{1-V^2/c^2}$ (where $V$ is the particle 
three-velocity), and from equation (\ref{pu}) 
(with $C_\alpha=0$) we find, 
\begin{equation}
\frac{\sqrt{2}C_v}{mc} = \sqrt{\frac{1 - \frac{V}{c}}{1 + \frac{V}{c}}}\,.
\label{cviav}
\end{equation}
One can then rewrite equation (\ref{tc1}) in a form displaying explicitly 
the dependence between the particle three-velocity $V$ and the angles 
$\theta_0$ and $\varphi_0$ as, 
\begin{equation}
\frac{1 - \frac{V}{c}}{1 + \frac{V}{c}} +
\cot^2{\frac{\theta_0}{2}}
\left(\cosh{2\beta} - \cos{2\varphi_0} \sinh{2\beta} \right) + 
 s^{(A)} q L^2 R^3_{\cdot u3u} = 0\,. 
\label{tc20}
\end{equation}
As it was mentioned at the end of last section, for $\theta_0=0$, 
i.e., the particle and the GW propagate in the same 
direction, the necessary condition for the emission of Cherenkov's radiation 
is not satisfied. 

Thus, we have to consider the particle moving in an opposite direction to
the GW, i.e. $\theta_0=\pi$.
Since $V<c$ we have from equation (\ref{tc20}) that the Cherenkov radiation 
exists when the following inequality holds, 
\begin{equation}
\cot^2{\frac{\theta_0}{2}}
\left(\cosh{2\beta} - \cos{2\varphi_0} \sinh{2\beta} \right) <
- s^{(A)} q L^2 R^3_{\cdot u3u}\,.
\label{tc2}
\end{equation}
This inequality (\ref{tc2}) describes the interior of the limit cone
obtained by puting $V=c$ in equation (\ref{tc20}), i.e., 
\begin{equation}
\cot^2{\frac{\theta_0}{2}}
\left(\cosh{2\beta} - \cos{2\varphi_0} \sinh{2\beta} \right) = 
- s^{(A)} q L^2 R^3_{\cdot u3u}\,.
\label{tc2equal}
\end{equation}
This equation defines a non-circular cone, depending not only on the 
azimuthal angle $\varphi_0$ but also on the retarded time $u$, 
via $\beta(u)$ and $L(u)$. The central axis of this cone coincides with
the direction $\theta_0 = \pi$. When $\beta(u)$ is negative, the instantaneous 
aperture of this cone is maximal at $\varphi_0 = \frac{\pi}{2}$, and is
minimal for $\varphi_0 = 0$. 

Looking at equation (\ref{tc2}) we find that the left-hand side 
of the inequality is always positive. Therefore, a necessary condition 
for the Cherenkov radiation is that $s^{(A)} q R^3_{\cdot
u3u}$ be negative. This cannot happen for both polarizations at 
the same time, but may happen alternately for each polarization. 
We deal here with an interesting phenomenon within the Cherenkov radiation, 
namely, polarized radiation with oscillating polarization direction.

\subsection{Transversal particle motion}

Let us consider now the following initial data for the particle motion, 
\begin{equation}
C_2 =0, 
\quad C_3 = - \frac{m V}{\sqrt{1-\frac{V^2}{c^2}}},
\quad C_v =  \frac{m c}{\sqrt2 \sqrt{1-\frac{V^2}{c^2}}}\,,
\label{transinitial}
\end{equation}
corresponding to transversal motion to the direction of the GW propagation, i.e., 
motion of the particle with three-velocity $V$ in $x^3$ direction 
on the wave front.
The criterion (\ref{gwcriterion}) gives in terms of $\theta_0$, $\varphi_0$,
and $V$ the following equation
\begin{eqnarray}
&
\frac{1}{L^2} \cot^2{\frac{\theta_0}{2}}
\left(\cosh{2\beta} - \cos{2\varphi_0} \sinh{2\beta} \right) +
s^{(A)} q R^3_{\cdot u3u} = 
\nonumber \\&
= -1 + \frac{V^2}{c^2} - \frac{V}{c} \frac{1}{L^2} e^{2\beta} 
\left(\frac{V}{c} - 2 \cot{\frac{\theta_0}{2}} \sin{\varphi_0} \right).
\label{transcriterion}
\end{eqnarray}
For a given $V$, this equation describes the cone along which the Cherenkov radiation 
can propagate. Since it depends on $\varphi_0$ it is a non-circular
cone. The aperture depends on the retarded time $u$, as in the previous case. 
The main axis of this cone is characterized by the  
polar angle $\pi/2$ and azimuthal angle $\pi/2$ (i.e., the $x^3-$direction). 
For an ultrarelativistic particle $V \to c$, and  in the section 
$\varphi_0 = \frac{\pi}{2}$,  equation (\ref{transcriterion}) can be reduced to
\begin{equation}
\left( \cot{\frac{\theta_0}{2}} - 1 \right)^2 = - 
s^{(A)} q R^3_{\cdot u3u} L^2 e^{- 2\beta}\,.
\label{pi2}
\end{equation}
Thus, the angle of emission of the created photon is predetermined by
the quantity 
$s^{(A)} q R^3_{\cdot u3u} L^2 e^{- 2\beta}$. Since this quantity is 
rather small, we deduce that this angle is very near to the direction 
of the particle motion given by the angle $\pi/2$.
Of course, Cherenkov radiation exists, when $s^{(A)} q R^3_{\cdot u3u}$ is 
negative, and this is possible for one of the two waves only.

\section{Conclusions}

\indent

We have seen that although a plane gravitational wave modifies the dielectric
properties of pure vacuum, it does not modify its scalar refraction index,
and therefore there is no possibility of emission of Cherenkov radiation
in this particular case.

On the other hand, vacuum interacting with curvature, considered as a sort of
quasi-medium, allows the possibility of the existence of Cherenkov
radiation.  We have shown that Cherenkov radiation in a 
vacuum interacting with curvature
can propagate along a noncircular cone, the spatial structure
of this cone being permanently modified with time. In addition, the Cherenkov
radiation exists alternatively for each polarization of the
electromagnetic wave. 

The dispersion relation for vacuum interacting with curvature is shown to be
analogous to a spatially anisotropic nonstationary medium. This analogy
allows one to explain the appearance of the Cherenkov radiation in the GW field
in terms of a transient radiation, first introduced by Ginzburg.

We have shown  a new effect, namely, the existence of 
polarized radiation with oscillating polarization direction, where the Cherenkov
angle is predetermined by the value of the Riemann tensor.

We have performed an exact treatment in two particular cases, for two
privileged directions of the particle motion, the longitudinal
and the transversal ones.  For arbitrary direction of particle
motion, the structure and the inclination of the cone's  axis with respect
to the GW front plane is given by more complicated modified expressions.

Now, we have used, in section 2, Ginzburg's way of explaining physically 
the origin of the Cherenkov radiation for a medium with constant
refraction index $n$.
However, as was shown in subsubsection 3.3.2. (see, e.g., the
formula (\ref{noku})), the effective refraction index for a vacuum interacting with
curvature is a function of the retarded time $u$. Since $n$ is not 
a constant one can then ask whether 
the
conclusion about Cherenkov radiation remains valid or not. 
The answer is positive.
To confirm  this conclusion one has to analyze equations 
(\ref{tidedalamber}) supplemented
by the electric current produced by the relativistic
particle. One can then obtain an exact solution
of the corresponding modified Maxwell equation and study it 
along the line of equations (\ref{threemax})-(\ref{oscillator}).
We have performed a preliminary calculation that fully supports this
conclusion. This preliminary calculation also shows that in the framework
of the
geometrical optics approximation, when the wavelength of GW is much larger
than the wavelength of the induced electromagnetic wave, we recover the
results we have obtained here for the structure of the Cherenkov light-cone. 
We hope to discuss
this problem in detail in a forthcoming paper.

\vskip 1cm
\indent
{\tbf Acknowledgments}
\vskip 0.2cm
\indent
This work has been supported by CNRS (France) during the invited stay of A.B.
and of J.P.S.L. in the Laboratoire G.C.R. at the University Paris-VI.
A.B. expresses his thanks to the CENTRA/IST in Lisbon for its hospitality,
and acknowledges a special grant for invited scientists from the Funda\c c\~ao
para a Ci\^encia e Tecnologia, through an ESO programme.

\newpage

{\noindent\Large\bf Appendix A. Solutions of Maxwell's equations in the GW
background}

\vskip 0.3cm

{\noindent\large\bf A1. Exact solution of Maxwell equations
for vacuum in the GW background}
\vskip 0.3cm

{\noindent\bf A1.1 Pure vacuum}

Let us recall briefly the main steps needed for solving Maxwell's equations 
\begin{equation}
\nabla_{k} H^{ik} = 0 \,,
\label{maxwell1}
\end{equation}
\begin{equation}
\nabla_{k} F^{*ik} =0 \,,
\label{maxwell2}
\end{equation}
where $F^{ik}$ is the Maxwell tensor,  $F^{*ik}$ its dual, and $H^{ik}$ is 
the induction tensor.
The set of equations (\ref{maxwell1}) and (\ref{maxwell2}) of covariant 
electrodynamics of continuous
media should be completed by a consistent formulation of the constitutive
equations \cite{maugin}, relating the tensor of electric and magnetic 
induction $H^{ik}$ with the Maxwell tensor.
The simplest relation between these tensors is linear and has the following
form \cite{maugin}:
\begin{equation}
H^{ik} = C^{ikmn}F_{mn} \,,
\label{constitutive}
\end{equation}
where $C^{ikmn}$ is the material tensor, describing the properties of 
linear response and containing the information about dielectric and magnetic
permeabilities, as well as about the magneto-electric coefficients.
In order to solve Maxwell's equations it is better to find a solution for the 
four-vector electromagnetic potential $A^i$, defined as 
\begin{equation}
F_{ik} = \nabla_i A_k - \nabla_k A_i \,.
\label{potentialdefinition}
\end{equation}
Working in the Lorentz gauge, 
\begin{equation}
\nabla_{k} \nabla^{k} A^{i} = 0\,,
\label{dalamber}
\end{equation}
then projecting the four-vector $A_i$ into the Killing directions,  
$\xi^{i}_{(v)}, \xi^{i}_{(2)}, \xi^{i}_{(3)}$, and after some 
manipulation, one finds the following solution (see \cite{bala1} 
for details), 
\begin{equation}
A_{(2)} =  e^{\beta} \ B_{(2)}(W)\,, \quad
A_{(3)} =  e^{- \beta} \ B_{(3)}(W)\,,
\label{a2a3sol}
\end{equation}
where $B_{(2)}(W)$ and $B_{(3)}(W)$ are arbitrary functions of the
argument $W$, which in turn is given by 
\begin{equation}
W = k_{v} \ v + k_{2} \ x^{2} +  k_{3} \ x^{3}  + \Phi (u) \,.
\label{phase}
\end{equation}
Here $  k_{v},  k_{2},  k_{3}$ are arbitrary constants, and $\Phi  (u)$
is obtained from the equation:
\begin{equation}
2k_{v} \Phi^{'}(u) + g^{\alpha \beta} k_{\alpha}  k_{\beta}  = 0\,.
\label{eqphi}
\end{equation}
When $k_{v} \not= 0$, we find
\begin{equation}
\Phi(u) = \Phi(0) - \frac{k_{\alpha}k_{\beta}}{2k_{v}}
\int^{u}_{o} du \  g^{\alpha \beta}(u)\,.  
\label{varphase}
\end{equation} 
The component $A_u$ is equal to
\begin{equation}
A_{u} = \frac{1}{k_{v}L^{2}}\left[e^{-\beta} \ k_{2} \ 
B_{(2)}(W) +  e^{\beta} \  k_{3} \  B_{(3)}(W) \right] \,.
\label{ausolution}
\end{equation}
If $k_v \equiv 0$, we obtain from (\ref{eqphi}) $k_2 = k_3 =0$ and the
function $\Phi(u)$ is totally arbitrary. The component $A_v$ does not enter
the Maxwell tensor and can be thus set to zero.

\vskip 0.3cm

{\noindent\bf A1.2 Vacuum interacting with curvature}
\vskip 0.3cm

Taking into account now QED corrections
\cite{drummond}, one can find that the induction and Maxwell tensors
are connected by the relationship $H^{ik} = F^{ik} + q R^{ikmn}
F_{mn}$, where $q= \alpha \lambda_e^2/90 \pi$ ($\alpha$ is the fine 
structure constant, and $\lambda_e$ is the Compton wavelength of the electron). 
Thus, we can rewrite the Maxwell equations in the form:
\begin{equation}
\nabla_{k} \nabla^{k} A^{i} = q \ R^{ikmn} \ \nabla_{k}
\left[\nabla_{m} A_{n} - \nabla_{n} A_{m} \right].
\label{tidedalamber}
\end{equation}
This is the appropriate quantum field theoretical modification of the
equation (\ref{dalamber}).

The solutions of Maxwell equation are given in this case by \cite{bala1}
\begin{equation}
A_{(2)} =  e^{\beta} \ B_{(2)}(W_{(2)})\,\quad 
A_{(3)} =  e^{- \beta} \ B_{(3)}(W_{(3)})\,,
\label{tidal23solution}
\end{equation}
where the phases $W_{(2)}$ and $W_{(3)}$ are now different from each other, 
\begin{equation}
W_{(2)} = W + k_{v} q \int^{u}_{0}d\tau \ R^{3}_{\cdot u3u}(\tau)\,,\quad
W_{(3)} = W - k_{v} q \int^{u}_{0}d\tau \ R^{3}_{\cdot u3u}(\tau)\,,
\label{phase23}
\end{equation}
and the $W$ scalar is the same as in (\ref{phase}). 
The $A_u$ component is now given by 
\begin{equation}
A_{u} = \frac{1}{k_{v}L^{2}}\left[e^{-\beta} \ k_{2} \ 
B_{(2)}(W_{(2)}) +  e^{\beta} \ k_{3} \ B_{(3)}(W_{(3)}) \right]\,.
\label{tidalausolution}
\end{equation}
The component $A_{(v)}$ does not contain any curvature contribution. Therefore,
it coincides with the equation for pure vacuum, and again we can put $A_{(v)} = 0$.

As an addendum, note that if $k_v = 0$, an arbitrary  function  of  the  
retarded time satisfies the equations (\ref{tidedalamber}) automatically. 
In other words, the electromagnetic waves  propagate in the same direction
as the gravitational wave, and ignore the interactions with curvature.

\vskip 0.5cm

{\noindent\large\bf A2. Solutions of Maxwell's equations in a spatially
isotropic medium  (electromagnetic waves in the geometrical optics 
approximation) }  
\vskip 0.3cm

The set of equations (\ref{maxwell1}) and (\ref{maxwell2}) of covariant 
electrodynamics of continuous media with linear constitutive equation 
(\ref{constitutive}) can be specified for the model under consideration in
the following way. When the medium, prior to the GW appearance, is 
isotropic and homogeneous, the tensor $C^{iklm}$ has the following structure
\cite{bala2}
\begin{equation}
C^{ikmn} = C^{ikmn}_{(\rm isotr)} + C^{ikmn}_{(\rm anisotr)} \,.
\label{csum}
\end{equation}
The standard isotropic part $C^{ikmn}_{(\rm isotr)}$ of the tensor $C^{ikmn}$
is given by
\begin{eqnarray}
&
C^{ikmn}_{(\rm isotr)} \equiv 
\frac{1}{2\mu} \left(g^{im} g^{kn} -
g^{in} g^{km} \right) +
\nonumber \\&
+ \left(\frac{\varepsilon \mu -1}{2\mu} \right)
\left(g^{im} U^k U^n - g^{in} U^k U^m + g^{kn} U^i U^m -
g^{km} U^i U^n \right) \,, 
\label{cisotr}
\end{eqnarray}
with dielectric permeability $\varepsilon$ and magnetic permeability $\mu$,
and the anisotropic contributions, linear in the Riemann tensor, are given by 
\begin{eqnarray}
&
C^{ikmn}_{(\rm aniso)} = 
Q R^{ikmn} +
\nonumber \\&
+ \bar{Q} U_p U_q \left(R^{ipmq} g^{kn} - R^{ipnq} g^{km} +
R^{kpnq} g^{im} - R^{kpmq} g^{in} \right) +
\nonumber \\&
+ \hat{Q} U_p U_q \left(R^{ipmq} U^k U^n - R^{ipnq} U^k U^m +
R^{kpnq} U^i U^m - R^{kpmq} U^i U^n \right) \,.
\label{caniso}
\end{eqnarray}
In the leading order approximation of geometrical optics \cite{syng} it is 
a common practice to represent the vector potential through a phase 
scalar $\Psi$ and a slowly varying amplitude $a_j$ defined by, 
\begin{equation}
A_j = a_j e^{i \Psi}, \quad K_j \equiv \nabla_j \Psi\,,
\label{geomoptics}
\end{equation}
where, $K_j$ is a wave-four vector. Assuming that the derivative of the phase
is much bigger than the derivative of the amplitude, we obtain for the Maxwell
tensor
\begin{equation}
F_{mn} = i(K_m a_n - K_n a_m) e^{i \Psi} \,.
\label{gomaxwelltensor}
\end{equation}
The Maxwell equation can then be reduced to
\begin{equation}
K_l K_n C^{ilmn} a_m = 0 \,.
\label{gomaxwellequation}
\end{equation}
Let us now consider the equation (\ref{gomaxwellequation}) in the case of
pure spatially isotropic medium, and then in the case of a medium interacting
with curvature.

\vskip 0.3cm

{\noindent\bf A2.1 Pure spatially isotropic medium }  
\vskip 0.3cm

In the absence of curvature induced effects, it is convenient to
introduce the generalized Lorentz condition
\begin{equation}
\nabla_j \left( A^j + (\varepsilon \mu - 1) U^j U_l A^l \right) = 0 \,,
\label{generalgauge}
\end{equation}
which coincides with standard Lorentz condition in vacuum 
($\varepsilon = \mu =1$), and when the vector potential $A_i$ is orthogonal
to the velocity four-vector $U^i$. In the eikonal approximation
equation (\ref{generalgauge}) yields
\begin{equation}
K_m a^m = (1 - \varepsilon \mu) K_l U^l a_n U^n \,.
\label{ka}
\end{equation}
Substituting the expression (\ref{cisotr}) 
into equation (\ref{gomaxwellequation}), one obtains, from equation (\ref{ka}):
\begin{equation}
\left(a^i + U^i (\varepsilon \mu - 1) a_m U^m \right)
\left[ K_m K^m + (\varepsilon \mu -1) (K_m U^m)^2 \right] = 0 \,.
\label{dispers1}
\end{equation}
A nontrivial $a^i$ vector exists if and only if the following relation takes
place, 
\begin{equation}
K_m K^m = (1 - \varepsilon \mu) (K_m U^m)^2 \,.
\label{dispers}
\end{equation}
So, comparing (\ref{dispers}) and (\ref{ksquare}), we can conclude,
that the index of refraction $n$ coincides identically with the index
of refraction of the medium $\sqrt{\varepsilon \mu}$, i.e., the
gravitational wave field does not modify the index of refraction of a
spatially isotropic medium.  Cherenkov radiation is possible for a
supraluminal moving particle, since for $\varepsilon \mu > 1$ the
necessary condition is satisfied, with or without a 
gravitational wave field.

\vskip 1.3cm

{\noindent\bf A2.2. Spatially isotropic medium  
interacting with curvature}  
\vskip 0.3cm

Let us consider the particular case, of a  transverse electromagnetic wave
propagating parallelely to the gravitational wave, i.e., $K_\alpha = 0$
and $a_u = a_v = 0$. As a consequence, we see, that $K_m a^m =0$.
Therefore, the generalized Lorentz condition (\ref{generalgauge}) leads 
to the relation $a_m U^m = 0$.

Taking into account formula (\ref{ka}), one can 
transform equation (\ref{gomaxwellequation}) with $C^{iklm}$ given by 
(\ref{csum})-(\ref{caniso}) into the following one
\begin{eqnarray}
&
a^i \left[ K_m K^m + (\varepsilon \mu -1) (K_m U^m)^2 \right] =
&\nonumber \\&
- \mu a^\alpha R^i_{\cdot u \alpha u}
\left[ 2 Q (K_v)^2  + \bar{Q} K_m K^m  + \hat{Q}(K_m U^m)^2  \right]\,.&
\label{tidaldispers1}
\end{eqnarray}
It includes three variables $K_v$, $K_m K^m$ and $K_m U^m$. Using the 
relationships
\begin{equation}
2 K_u K_v = K_m K^m \,, \quad \frac{1}{\sqrt{2}}(K_u + K_v) = K_m U^m\,,
\label{rela}
\end{equation}
one can express the first variable $K_v$  in terms of the scalar index of 
refraction $n$, 
and the other two variables, $K_m K^m$ and $K_m U^m$, through the relation, 
\begin{equation}
K_v = \frac{1}{\sqrt{2}} K_m U^m \left(
1 \pm \sqrt{1 - \frac{K_m K^m}{(K_mU^m)^2}} \right) =
\frac{1}{\sqrt{2}} K_m U^m (1 \pm n) \,.
\label{kv}
\end{equation}
Considering equation (\ref{tidaldispers1}) for $i = u, v$, we obtain 
a trivial equality. The nontrivial solutions for $a^{(A)}$ ($(A) = 2,3)$
do exist when  $n = n^{(A)}$ satisfies the following quadratic equation,
\begin{eqnarray}
&
(n^{(A)})^2 \left[1 - \mu s^{(A)} R^3_{\cdot u3u}(\bar{Q} - Q) \right]
\pm 2 n^{(A)} \mu s^{(A)} R^3_{\cdot u3u} Q -
\nonumber \\&
- \mu \left[\varepsilon  -  s^{(A)} R^3_{\cdot u3u}(Q + \bar{Q} + \hat{Q})
\right] = 0 \,.
\label{quadraticequation}
\end{eqnarray}
The medium becomes anisotropic due to curvature interactions, the  index 
of refraction depends on the polarization and, consequently, 
birefringence  appears in the same way as in the vacuum case.

\vskip 1cm

{\noindent\Large\bf Appendix B. Another case admiting Cherenkov
radiation: Spatially isotropic medium interacting \hfill\break 
with curvature}
\vskip 0.3cm

As it is shown in Appendix A (see section A2.2), birefringence appears in the case
of vacuum interacting with curvature. We obtain for the indices of refraction
the following expressions :
\begin{eqnarray}
&
n^{(A)} =  \left[1 - \mu s^{(A)} R^3_{\cdot u3u}(\bar{Q} - Q) \right]^{-1}
\left\{\mp \mu s^{(A)} R^3_{\cdot u3u} Q  + 
\right.
\nonumber \\&
\left.
+ \sqrt{ \mu^2 (R^3_{\cdot u3u})^2 Q^2 + 
\mu \left[\varepsilon  -  s^{(A)} R^3_{\cdot u3u}(Q + \bar{Q} + \hat{Q})
\right] 
\left[1 - \mu s^{(A)} R^3_{\cdot u3u}(\bar{Q} - Q) \right]} \right\}.
\label{quadraticequation0}
\end{eqnarray}
The sign plus in front of the square route was thus chosen because in the
absence of GW the solution (\ref{quadraticequation0}) have to coincide
with $\sqrt{\mu \varepsilon}$.  In the GW field the indices of
refraction $n^{(A)}$ can be in general larger and smaller than the
index of refraction in the medium $\sqrt{\varepsilon \mu}$.  The index
of refraction, being a function of the Riemann tensor component
$R^3_{\cdot u3u}$, changes periodically. Thus, at least for one of the
electromagnetic waves with orthogonal polarization the necessary
condition for the Cherenkov radiation is satisfied. When the scalars $\mu$
and $\varepsilon$ are close to one the value of Riemann tensor
predetermines the phenomenon. Depending on the case, one or both waves
can propagate with subluminal phase velocity.
Thus, either both waves or only one of them can describe a wave emitted
due to the Cherenkov effect.

\end{document}